# A VENDOR'S EVALUATION–USING AHP FOR AN INDIAN STEEL PIPE MANUFACTURING COMPANY

Giridhar Kamath
Rakesh Naik
Shiva Prasad H C
Manipal Institute of Technology
Department of Humanities and Management
Manipal University, Manipal
India
giridhar.bk@yahoo.com,
rakeshnr.16@gmail.com
hcs.prasad@manipal.edu

## ABSTRACT

To improve a firm's supply chain performance it is essential to have a vendor evaluation process to be able to showcase an organization's success in the present aggressive market. Hence, the process of evaluating the vendor is a crucial task of the purchasing executives in supply chain management. The objective of this research is to propose a methodology to evaluate the vendors for a steel pipe manufacturing firm in Gujarat, India. For the purpose of the study, the Analytical Hierarchy Process was used to evaluate the best raw material vendor for this company. Multiple qualitative and quantitative criteria are involved in the vendor evaluation process. To solve the complex problem of vendor evaluation, a tradeoff between these multi-criteria is important. The outcomes indicated that the AHP technique makes it simpler to assign weights for the different criteria for evaluating the vendor. Research findings showed that quality is the most important criterion followed by delivery, cost and vendor relationship management.

## 1. Introduction
The field of supply management has been undergoing a transformation from a tactical, transaction oriented role to a strategic capability at many companies. Senior executives are discovering that a good, integrated supply management capability is not only necessary, but also required to achieve a competitive advantage. Management is realizing that there is potential for procurement to add cash to the bottom line instead of only viewing procurement as a cost center. The procurement function that used to play a lesser role in organizations now has titles like chief procurement officer and corporate vice president of vendor quality and performance management. The senior level executives have to consider many factors such as goals of the organization, financial tolerance, administrative time frame and many more when evaluating a vendor. The raw material costs account for up to nearly 70% of the total cost of the product in many industries (Ghodsypour & O'Brien, 1998). Hence, an overall goal of cost reduction depends upon decisions made by the procurement executives. For the success of an organization in a competitive market, selection and evaluation of the vendor plays an important role.

For the buying firm, evaluating the best vendor is one of the most challenging tasks. The varied strengths and weaknesses of the vendors make it difficult for the purchasing firm to carefully assess the vendors before ranking them. The evaluation of the vendor would be simple if only one criterion was used in the process of making the decision. However, for the purchasing executives of a steel pipe manufacturing firm, evaluation of the vendor involves a number of criteria and sub-criteria. Therefore, if many criteria are involved in the process of selecting the best vendor it is essential to determine whether the criteria are equally weighted or vary according to the type of criteria (Yahya & Kingsman, 1999). The development of the model for XYZ company's vendor evaluation is essential not only for the benefit of the organization, but also because the vast range of the finished products requires a vast range of raw materials which fluctuate in price, and there are a large number of vendors and projects in process.

The vendors are evaluated based on a number of criteria so that the purchasing department can make their vendor selection decision based upon the most essential criteria. Evaluation of the vendor is a group Multiple Criteria Decision Making (MCDM) problem (Ho, Xu & Dey, 2010). In MCDM the purchasing executives have to analyze the tradeoff among several conflicting criteria in vendor evaluation. The Analytical Hierarchy Process (AHP) is a linear weighing model, and is one of the most used models among the various approaches for the evaluation (John, Baby & Mangalathu, 2013; Sonawane, & Rodrigues, 2015). The AHP technique is recommended for the criteria selection in vendor evaluation to ease or eliminate the inaccuracy in this line which is often caused by adjudicating the raw materials or giving attention to only one criterion such as cost or quality.

**Research objectives**
The main objective of the research is to identify the measure or criteria that impact the evaluation of the vendor at XYZ firm. Thus, the objectives are:
1. To determine the factors that influences the vendor evaluation at XYZ firm.
2. To develop a model that describes the measure for evaluating the vendor.
3. To determine the best raw material supply vendor.

**Research questions**
1. What are the criteria for vendor evaluation at XYZ firm?
2. Which is the best raw material vendor at XYZ firm?

## 2. Literature review
The process of vendor evaluation becomes a very complicated task as many criteria should be taken into account with more than 20 factors suggested for the evaluation of the vendor that the procurement managers have to consider during the process of selection of the vendor (Dickson, 1966; Imeri, 2013). The purchasing managers do a lot more than just buying goods. The main job of the managers is to make decisions regarding important criteria along with other people in the organization. Other than minimizing the cost, the responsibility of the managers of the procurement department is to select the appropriate vendor to help them accomplish the wide objectives of the firm.

While meeting the organization's goal, evaluation of the vendor process helps recognize and differentiate between vendors at an adequate cost. Based on the criteria considered for the vendor evaluation, a number of vendors are being selected. In order for an organization to remain sustainable in the competitive market they must provide better quality and services to their customers to satisfy them. Therefore, the company should evaluate and select the vendor that is best able to make sure they manufacture a quality product. In order to do this, the company has to spend a significant amount of time evaluating the suitable vendor (Alsuwehri, 2011).

It is essential to identify the criteria that influence the vendor evaluation process. As suggested by Dickson (1966) for vendor selection from a group of criteria, the important criteria like lowest price, quality, and prompt deliveries are considered by many researchers and common metrics used (Dickson, 1966; Weber, Current, & Benton,1991). Based on twenty-three criteria presented in the studies of Weber, Current, & Benton (1991) and Dickson (1966), the most important criteria for evaluation of the vendor are quality, price and delivery and management and organization (Alsuwehri, 2011). Based on the above literature review, quality, cost, delivery and vendor relationship management were the criteria considered for the organization in the case study. The sub-criteria under quality are specification of the raw material or the equipment, warranty, rejection, packing, continuous improvement and top management (Yusuff, Yee & Hashmi, 2001). Under the cost criteria net price, ordering and delivery cost, and capital investments are the sub-criteria (Yusuff, Yee & Hashmi, 2001). The measures of delivery for vendor evaluation are late delivery, location and lead time (Alsuwehri, 2011). Another criterion that is taken into consideration for the study is vendor relationship management (VRM). The process of building and maintaining a sustainable, cordial relation with the supplier with social fabrics is the basis of the relationship apart from the formal business transaction (Giunipero & Pearcy, 2000). Hence, managing the social fabric with the vendor is essential at the nascent stage of business process. Due to the commitment to multiple partnerships and since the relationship among the partners is dynamic (not everlasting), vendor relationship management is a crucial issue in the evaluation of the vendor in supply chain management (Giunipero & Pearcy, 2000). This is because substitution of the partners has become a common practice in the industry in order for companies to improve their performance and meet the multiple market conditions (Mowshowitz, 1997). The vendor's involvement in research and development activities helps the company with continuous improvement (Tahriri, et al., 2008). A long-term relationship with the vendor is crucial for a firm to be able to effectively fulfill the demands of the customers (Kannan & Tan, 2002). The reputation of the vendor should also be taken into account as another sub-criterion in vendor relationship management as it facilitates improvement in both the business and operational perspectives (Kannan & Tan, 2002).

Constant vendor evaluation, selection and measurement of performance are essential for the success of every organization. This is especially important when developing a new product. Vendor evaluation is a problem that involves multiple criteria which include both tangible and intangible factors. Therefore, for a company to acquire higher profits even a small cost reduction in procuring the raw material will make a difference. The process of vendor evaluation has a direct influence on quality, cost, delivery and vendor relationship management.

# 3. Methodology

The specific scheme discussed in the paper is for XYZ steel pipe manufacturing company in Gujarat, India. The main product of the company is welded mild steel (MS) pipe. The company also manufactures MS plates and coils, bend pipes and offers coating for the pipes. The pipe plants of XYZ Company are located in India, the U.S. and Saudi Arabia. The main goal of the research is to evaluate the best raw material vendor using a combination of the AHP technique and the factor rating method. Complex, multi-attribute problems can be handled effectively by the AHP technique (Yusuff, Yee & Hashmi, 2001). The AHP technique is found to be useful to help reach a consensus solution to a problem with diverse and conflicting criteria (Tam & Tummala, 2001). Hence, the AHP is found to be very effective in the problem of evaluating the vendor decision to determine the best vendor (Yu, & Jing, 2004). The AHP approach is used to assign weights to the various criteria, and the factor rating method is used to find the optimal vendor based on the criteria. The criteria and sub criteria for vendor evaluation were identified upon close interaction with the managers and the Deputy General Manager of the procurement, finance, and quality control departments (see Figure 1). The credibility of the research was not affected by this method as managers and the DGM had in-depth knowledge and experience with respect to the raw material vendors. Six raw material vendors, each providing American Petroleum Institute (API) (www.fedsteel.com) and Indian Standard (IS) (www.gipipesindia.com), specification raw material were considered for the study. To ensure the effective development of the model the following five steps were implemented.

## 3.1 Steps
**Step 1: State the criteria for vendor evaluation**
Establishing the criteria is the first step in the process of vendor evaluation. The criteria for vendor evaluation are quality, cost, delivery and vendor relationship management.

**Step 2: Define sub-criteria for vendor evaluation**
The second step is to define the sub-criteria for the above-mentioned criteria. The sub-criteria were selected based on the literature review and discussion with the managers and senior level executives of the purchasing, finance and quality control departments. A total of twenty-four sub-criteria were considered for the study; six sub-criteria under each criterion were considered. The sub-criteria that will result in the delivery of raw material by the vendors taking into account the most essential requirements of the organization were selected.

**Step 3: Structuring of hierarchy model**
In this step, weights were assigned to the criteria and sub-criteria. Weights were allocated for each criteria and sub-criteria by making pair-wise comparisons with the AHP technique. Pair-wise comparison was carried out by obtaining the relative importance for the criteria and sub-criteria. A nine-point scale depicted by Saaty (1980) was used for this. The nine-point scale specifies the comparative significance with the levels equal, moderate, strong, very strong, and extreme represented by 1, 3, 5, 7, and 9, respectively. The intermediary importance between two contiguous contrasts is indicated by 2, 4, 6, and 8. Experts have accepted the nine-point scale depicted by Saaty as it is very scientific for comparisons of two alternate criteria (Saaty, 1980). In the AHP technique, the fundamental assumption for the comparison of the criteria is; if criteria X is very strongly preferred to criteria Y then it is rated as 7, and if Y is strongly preferred to X it is rated as 1/7 (Saaty, 1980). The pair-wise comparison is carried out for all the criteria (see Table 1).

Table 1
Pair-wise comparison values

| Preferences | Ratings |
| --- | --- |
| Equally preferred | 1 |
| Equally to moderately preferred | 2 |
| Moderately preferred | 3 |
| Moderately to strongly preferred | 4 |
| Strongly preferred | 5 |
| Strongly to very strongly preferred | 6 |
| Very strongly preferred | 7 |
| Very strongly to extremely preferred | 8 |
| Extremely preferred | 9 |

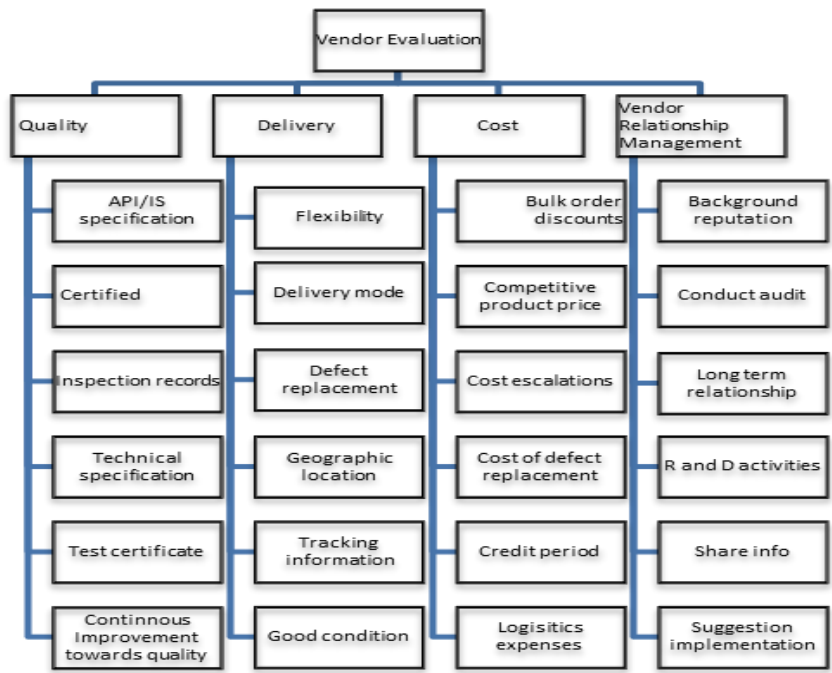

Figure 1. Criteria and sub-criteria for vendor evaluation

The objective of the study is to evaluate the vendors based on the criteria mentioned for XYZ Company. To assign the weights for the criteria, the following steps have to be followed (Saaty, 1980).

1. Find the sum of values in each column of the pair-wise comparison matrices.
2. Divide each value in each column by the parallel column sum. The outcome denotes the normalized matrices.
3. Calculate the average of each row of the normalized matrices. The outcome depicts preference vectors.
4. Combine the preference vectors for each criterion into one preference matrix that depicts the weights for each criterion.

**Consistency check**

The comparisons of the criteria made by the executives of the organization have to be validated for consistency to make sure that the model is reliable. The decision made by the executives is consistent, if the Consistency Ratio (CR) is zero. The CR value for the outcomes of the results is acceptable if the values are less than 0.1 as given by Saaty and Hu (1998). A CR value above 0.1 is unacceptable, and these results are considered untrustworthy because it is too close to randomness. In this case, the decision may have to be repeated (Saaty & Hu, 1998).

**Consistency Ratio (CR) calculation**

A CR is calculated by computing $\lambda_{max}$ which facilitates the calculation of Consistency Index (CI) taking into account Equation 1.

$$A_x = \lambda_{max} x \quad (1)$$

where A is the preference matrix and x is the eigenvector, so

$$\lambda \max = \text{average}\left(\frac{Ax}{x}\right) \quad (2)$$

CI is computed by the Equation 2.

$$CI = \frac{\lambda \max - n}{n - 1} \quad (3)$$

The equation for consistency ratio is given by

$$CR = \frac{CI}{RI} \quad (4)$$

where CI is the consistency index and RI is the index for the corresponding random matrix. The RI values are depicted from Saaty's table (Table 2).

Table 2
Reference value of RI (Saaty & Tran, 2007)

| Order of randomness | 1 | 2 | 3 | 4 | 5 | 6 | 7 | 8 | 9 | 10 |
|---|---|---|---|---|---|---|---|---|---|---|
| Random Index (RI) | 0.00 | 0.00 | 0.58 | 0.90 | 1.12 | 1.24 | 1.32 | 1.41 | 1.45 | 1.49 |

A pair-wise comparison study for criteria considered for vendor evaluation is shown in Table 3.

Table 3
Pair-wise comparison of criteria

| Criteria | QUALITY | COST | DELIVERY | VRM |
|---|---|---|---|---|
| QUALITY | 1 | 9 | 5 | 9 |
| COST | 1/9 | 1 | 1/4 | 3 |
| DELIVERY | 1/5 | 4 | 1 | 5 |
| VRM | 1/9 | 1/3 | 1/5 | 1 |

Table 4
Pair-wise comparison and column sums

| Criteria | QUALITY | COST | DELIVERY | VRM |
|---|---|---|---|---|
| QUALITY | 1 | 9 | 5 | 9 |
| COST | 0.111 | 1 | 0.25 | 3 |
| DELIVERY | 0.2 | 4 | 1 | 5 |
| VRM | 0.111 | 0.333 | 0.2 | 1 |
| Column Sums | 1.422 | 14.333 | 6.45 | 18 |

Table 5
Normalized column sums

| | QUALITY | COST | DELIVERY | VRM | Local weight (row averages) |
|---|---|---|---|---|---|
| QUALITY | 0.703 | 0.628 | 0.775 | 0.500 | 0.652 |
| COST | 0.078 | 0.070 | 0.039 | 0.167 | 0.088 |
| DELIVERY | 0.141 | 0.279 | 0.155 | 0.278 | 0.213 |
| VRM | 0.078 | 0.023 | 0.031 | 0.056 | 0.047 |

Thus, with pair-wise comparison it can be determined that the company extremely prefers quality over cost criterion, and therefore it has been rated nine as defined in Saaty's table (Table 1). The preference of cost over quality is therefore rated as 1/9 (0.1111) which is the inverse of the preference of quality over cost criterion. Similarly, the company strongly preferred quality over delivery and it is rated five, and quality is extremely preferred over vendor relationship management (VRM) represented with a nine. The company moderately prefers cost over vendor relationship management (rated as three). The company strongly to moderately prefers delivery over cost and strongly prefers over VRM. The criterion compared with itself is rated as one.

In order to determine whether the responses of the experts of the procurement department are consistent after the pair-wise comparisons it is necessary to find the consistency ratio (CR). This is computed as given with Equation 4. λmax was computed with Equation 2 and is shown in Table 6.

Table 6
λmax calculation

|   | A |   |   |   |   | X |   | AX |   | AX |
|---|---|---|---|---|---|---|---|---|---|---|
| 1 | 9 | 5 | 9 | * | 0.652 | = | 2.935 |   |   | 4.505 |
| 0.111 | 1 | 0.3 | 3 | * | 0.088 | = | 0.355 |   | $\lambda_{max}$ | 4.019 |
| 0.2 | 4 | 1 | 5 | * | 0.213 | = | 0.932 | = |   | 4.371 |
| 0.111 | 0.333 | 0.2 | 1 | * | 0.047 | = | 0.191 |   |   | 4.075 |

$\lambda_{max}$ = average (4.5047/2.9350, 4.0186/0.3550, 4.3715/0.9317, 4.0746/0.1914) = 4.2424

Consistency Index = $(\lambda_{max}-n)/(n-1)$;  i e, (4.2424-4)/ (4-1) = 0.0808

Consistency Ratio (CR) = CI/RI = 0.0808/0.9 = 0.0898<0.1; therefore, the pair-wise comparisons of criteria for vendor evaluation are consistent. The value of RI for 4 orders of randomness (4 criteria) is 0.9. The row averages from the normalized table determined that quality is the most important criterion with a weight of 65% followed by delivery (21%), cost (9%) and vendor relationship management (4%) for the XYZ firm. The pair-wise comparisons, normalized table and consistency check for the sub-criteria are analyzed in Tables 7 to 18 and the results are analyzed for a consistency check in Table 19.

Table 7
Pair-wise comparison between sub-criteria of quality criterion

| SC-quality | TS | I | S | C | TC | CI |
|---|---|---|---|---|---|---|
| TS | 1 | 1 | 1 | 1 | 1 | 5 |
| I | 1 | 1 | 1 | 1 | 1 | 6 |
| S | 1 | 1 | 1 | 1 | 1 | 8 |
| C | 1 | 1 | 1 | 1 | 1 | 8 |
| TC | 1 | 1 | 1 | 1 | 1 | 8 |
| CI | 0.2 | 0.167 | 0.125 | 0.125 | 0.125 | 1 |
| SUM | 5.2 | 5.167 | 5.125 | 5.125 | 5.125 | 36 |

TS-Technical specification, I-Inspections, S-API/ IS specification, C-Certified, TC-Test certificate, CI- Continuous improvement

Table 8
Normalized table of sub-criteria of quality

|   | TS | I | S | C | TC | CI | Local weights |
|---|---|---|---|---|---|---|---|
| TS | 0.192 | 0.194 | 0.195 | 0.195 | 0.195 | 0.139 | 0.185 |
| I | 0.192 | 0.194 | 0.195 | 0.195 | 0.195 | 0.167 | 0.190 |
| S | 0.192 | 0.194 | 0.195 | 0.195 | 0.195 | 0.222 | 0.199 |
| C | 0.192 | 0.194 | 0.195 | 0.195 | 0.195 | 0.222 | 0.199 |
| TC | 0.192 | 0.194 | 0.195 | 0.195 | 0.195 | 0.222 | 0.199 |
| CI | 0.038 | 0.032 | 0.024 | 0.024 | 0.024 | 0.028 | 0.029 |

Table 9
$\lambda$max calculation

| A | | | | | | | X | | AX | | | AX |
|---|---|---|---|---|---|---|---|---|---|---|---|---|
| 1 | 1 | 1 | 1 | 1 | 5 | * | 0.185 | = | 1.114 | | | 6.023 |
| 1 | 1 | 1 | 1 | 1 | 6 | * | 0.190 | = | 1.143 | | | 6.027 |
| 1 | 1 | 1 | 1 | 1 | 8 | * | 0.199 | = | 1.200 | | | 6.034 |
| 1 | 1 | 1 | 1 | 1 | 8 | * | 0.199 | = | 1.200 | = | $\lambda_{max}$ | 6.034 |
| 1 | 1 | 1 | 1 | 1 | 8 | * | 0.199 | = | 1.200 | | | 6.034 |
| 0.200 | 0.167 | 0.125 | 0.125 | 0.125 | 1 | * | 0.0299 | = | 0.172 | | | 6.005 |

Table 10
Pair-wise comparison between sub-criteria of cost criterion

| SC-cost | L | BOD | DRC | CE | CP | CPP |
|---|---|---|---|---|---|---|
| L | 1 | 0.143 | 1 | 1 | 0.111 | 0.111 |
| BOD | 7 | 1 | 1 | 5 | 0.143 | 0.2 |
| DRC | 1 | 1 | 1 | 1 | 0.125 | 0.125 |
| CE | 1 | 0.2 | 1 | 1 | 0.125 | 0.143 |
| CP | 9 | 7 | 8 | 8 | 1 | 1 |
| CPP | 9 | 5 | 8 | 7 | 1 | 1 |
| SUM | 28 | 14.343 | 20 | 23 | 2.504 | 2.579 |

L-Logistics, BOD-Bulk Order Discounts, DRC- Defect Replacement Cost, CE-Cost Escalations, CP-Credit Period, CPP-Competitive Product Price

Table 11
Normalized table of sub-criteria of cost

| | L | BOD | DRC | CE | CP | CPP | Local weights |
|---|---|---|---|---|---|---|---|
| L | 0.036 | 0.010 | 0.050 | 0.043 | 0.044 | 0.043 | 0.038 |
| BOD | 0.250 | 0.070 | 0.050 | 0.217 | 0.057 | 0.078 | 0.120 |
| DRC | 0.036 | 0.070 | 0.050 | 0.043 | 0.050 | 0.048 | 0.050 |
| CE | 0.036 | 0.014 | 0.050 | 0.043 | 0.050 | 0.055 | 0.041 |
| CP | 0.321 | 0.488 | 0.400 | 0.348 | 0.399 | 0.388 | 0.391 |
| CPP | 0.321 | 0.349 | 0.400 | 0.304 | 0.399 | 0.388 | 0.360 |

Table 12
$\lambda_{max}$ calculation

| A | | | | | | | X | | AX | | | AX |
|---|---|---|---|---|---|---|---|---|---|---|---|---|
| 1 | 0.143 | 1 | 1 | 0.111 | 0.111 | * | 0.038 | = | 0.229 | | | 6.073 |
| 7 | 1 | 1 | 5 | 0.143 | 0.200 | * | 0.120 | = | 0.769 | | | 6.394 |
| 1 | 1 | 1 | 1 | 0.125 | 0.125 | * | 0.050 | = | 0.343 | | | 6.920 |
| 1 | 0.2 | 1 | 1 | 0.125 | 0.143 | * | 0.041 | = | 0.253 | = | $\lambda_{max}$ | 6.112 |
| 9 | 7 | 8 | 8 | 1 | 1 | * | 0.391 | = | 2.661 | | | 6.809 |
| 9 | 5 | 8 | 7 | 1 | 1 | * | 0.360 | = | 2.379 | | | 6.603 |

Table 13
Pair-wise comparison between sub-criteria of delivery criterion

|     | F     | TI  | GL    | DM   | GC    | DRT |
|-----|-------|-----|-------|------|-------|-----|
| F   | 1     | 8   | 1     | 5    | 1     | 1   |
| TI  | 0.125 | 1   | 0.2   | 0.2  | 0.125 | 0.2 |
| GL  | 1     | 5   | 1     | 7    | 1     | 1   |
| DM  | 0.2   | 5   | 0.143 | 1    | 0.143 | 0.2 |
| GC  | 1     | 8   | 1     | 7    | 1     | 5   |
| DRT | 1     | 5   | 1     | 5    | 0.2   | 1   |
| SUM | 4.325 | 32  | 4.343 | 25.2 | 3.468 | 8.4 |

F-Flexibility, TI-Tracking Information, GL-Geographic Location, DM-Delivery Mode, GC-Good Condition, DRT-Defect Replacement Time

Table 14
Normalized column of sub-criteria of delivery

|     | F     | TI    | GL    | DM    | GC    | DRT   | Local weights |
|-----|-------|-------|-------|-------|-------|-------|---------------|
| F   | 0.231 | 0.250 | 0.230 | 0.198 | 0.288 | 0.119 | 0.220         |
| TI  | 0.029 | 0.031 | 0.046 | 0.008 | 0.036 | 0.024 | 0.029         |
| GL  | 0.231 | 0.156 | 0.230 | 0.278 | 0.288 | 0.119 | 0.217         |
| DM  | 0.046 | 0.156 | 0.033 | 0.040 | 0.041 | 0.024 | 0.057         |
| GC  | 0.231 | 0.250 | 0.230 | 0.278 | 0.288 | 0.595 | 0.312         |
| DRT | 0.231 | 0.156 | 0.230 | 0.198 | 0.058 | 0.119 | 0.165         |

Table 15
$\lambda_{max}$ calculation

| A   |   |       |   |       |     |   | X     |   | AX    |   |               | AX    |
|-----|---|-------|---|-------|-----|---|-------|---|-------|---|---------------|-------|
| 1   | 8 | 1     | 5 | 1     | 1   | * | 0.220 | = | 1.430 |   |               | 6.512 |
| 0.125 | 1 | 0.2 | 0.2 | 0.125 | 0.2 | * | 0.029 | = | 0.183 |   |               | 6.322 |
| 1   | 5 | 1     | 7 | 1     | 1   | * | 0.217 | = | 1.456 | = | $\lambda_{max}$ | 6.705 |
| 0.2 | 5 | 0.143 | 1 | 0.143 | 0.2 | * | 0.057 | = | 0.354 |   |               | 6.251 |
| 1   | 8 | 1     | 7 | 1     | 1   | * | 0.312 | = | 2.205 |   |               | 7.064 |
| 1   | 5 | 1     | 5 | 0.2   | 0.2 | * | 0.165 | = | 1.093 |   |               | 6.605 |

Table 16
Pair-wise comparison between sub-criteria of vendor relationship management

|     | CA | SI | BR | LR | RDA | SUI |
|-----|------|------|------|------|------|------|
| CA  | 1    | 1    | 0.20 | 0.20 | 1    | 1    |
| SI  | 1    | 1    | 0.20 | 0.20 | 1    | 1    |
| BR  | 5    | 5    | 1    | 1    | 6    | 1    |
| LR  | 5    | 5    | 1    | 1    | 7    | 6    |
| RDA | 1    | 1    | 0.17 | 0.143| 1    | 1    |
| SUI | 1    | 1    | 1    | 0.167| 1    | 1    |
| Sum | 14.00| 14.00| 3.57 | 2.71 | 17.00| 11.00|

CA-Conduct Audit, SI-Share Information, BR-Background Reputation, LR-Long-term Relationship, RDA-R & D Activities, SUI-Suggestion Implementation

Table 17
Normalized table of sub-criteria of vendor relationship management

|     | CA    | SI    | BR    | LR    | RDA   | SUI   | Local weights |
|-----|-------|-------|-------|-------|-------|-------|---------------|
| CA  | 0.071 | 0.071 | 0.056 | 0.074 | 0.059 | 0.091 | 0.070         |
| SI  | 0.071 | 0.071 | 0.056 | 0.074 | 0.059 | 0.091 | 0.070         |
| BR  | 0.357 | 0.357 | 0.280 | 0.369 | 0.353 | 0.091 | 0.301         |
| LR  | 0.357 | 0.357 | 0.280 | 0.369 | 0.412 | 0.545 | 0.387         |
| RDA | 0.071 | 0.071 | 0.047 | 0.053 | 0.059 | 0.091 | 0.065         |
| SUI | 0.071 | 0.071 | 0.280 | 0.062 | 0.059 | 0.091 | 0.106         |

Table 18
$\lambda_{max}$ calculation

| A | | | | | | | X | | AX | | AX |
|---|---|---|---|---|---|---|---|---|---|---|---|
| 1 | 8 | 1 | 5 | 1 | 1 | * | 0.220 | = | 1.430 | | 6.512 |
| 0.125 | 1 | 0.2 | 0.2 | 0.125 | 0.2 | * | 0.029 | = | 0.183 | | 6.322 |
| 1 | 5 | 1 | 7 | 1 | 1 | * | 0.217 | = | 1.456 | $= \lambda_{max}$ | 6.705 |
| 0.2 | 5 | 0.143 | 1 | 0.143 | 0.2 | * | 0.057 | = | 0.354 | | 6.251 |
| 1 | 8 | 1 | 7 | 1 | 1 | * | 0.312 | = | 2.205 | | 7.064 |
| 1 | 5 | 1 | 5 | 0.2 | 0.2 | * | 0.165 | = | 1.093 | | 6.605 |

From Equations 2, 3 and 4 a consistency check for the sub-criteria of quality, cost, delivery and VRM is computed and shown in Table 19.

Table 19
Consistency check

|                          | λmax  | CI    | CR    | Decision   |
|--------------------------|-------|-------|-------|------------|
| Sub-criteria of Quality  | 6.026 | 0.005 | 0.004 | consistent |
| Sub-criteria of Cost     | 6.576 | 0.115 | 0.093 | consistent |
| Sub-criteria of Delivery | 6.485 | 0.097 | 0.078 | consistent |
| Sub-criteria of VRM      | 6.377 | 0.075 | 0.061 | consistent |

The global weights for each sub-criteria are computed by the product of local weights of sub-criteria and its relevant criteria and shown in Table 20. Quality is the most important criteria with the highest local weight of 0.652 followed by delivery (0.213), cost (0.088) and VRM (0.047) (Table 20). The prioritization of sub-criteria depends on the local weights

Table 20
Assignment of local and global weights

| Criteria | Local weight | Sub criteria | Local weights | Global weights | Criteria | Local weight | Sub criteria | Local weights | Global weights |
|---|---|---|---|---|---|---|---|---|---|
| Quality | 0.652 | API/ IS spec | 0.199 | 0.130 | Delivery | 0.213 | Good condition | 0.312 | 0.035 |
| | | Certified | 0.199 | 0.130 | | | Flexibility | 0.220 | 0.032 |
| | | Test certificate | 0.199 | 0.130 | | | Geographic location | 0.217 | 0.011 |
| | | Inspection | 0.190 | 0.124 | | | Defect replacement | 0.165 | 0.004 |
| | | Technical specification | 0.185 | 0.121 | | | Delivery mode | 0.057 | 0.004 |
| | | Continuous improvement | 0.029 | 0.019 | | | Tracking information | 0.029 | 0.003 |
| Cost | 0.088 | Credit period | 0.391 | 0.067 | VRM | 0.047 | Long term relationship | 0.387 | 0.018 |
| | | Competitive product price | 0.360 | 0.047 | | | Background reputation | 0.301 | 0.014 |
| | | Bulk order discounts | 0.120 | 0.046 | | | Suggestion implementation | 0.106 | 0.005 |
| | | Cost of defect replacement | 0.050 | 0.035 | | | Conduct audit | 0.070 | 0.003 |
| | | Cost escalations | 0.041 | 0.012 | | | Share info | 0.070 | 0.003 |
| | | Logistics cost | 0.038 | 0.006 | | | R and D activities | 0.065 | 0.003 |

**Step 4: Sub-criteria order prioritization**
After the completion of pair-wise comparisons, and calculating the local weights for each criterion, the next step is to arrange the criteria according to the level of importance for evaluating the best vendor.

Table 21
Prioritization of global weights

| Sub criteria | Global weights | Weights in % | Sub criteria | Global weights | weights in % |
|---|---|---|---|---|---|
| API/ IS spec | 0.130 | 13.0 | Long term relationship | 0.018 | 1.8 |
| Certified | 0.130 | 13.0 | Background reputation | 0.014 | 1.4 |
| Test certificate | 0.130 | 13.0 | Delivery mode | 0.012 | 1.2 |
| Inspection | 0.124 | 12.4 | Bulk order discounts | 0.011 | 1.1 |
| Technical specification | 0.121 | 12.1 | Tracking information | 0.006 | 0.6 |
| Good condition | 0.067 | 6.7 | Suggestion implementation | 0.005 | 0.5 |
| Flexibility | 0.047 | 4.7 | Cost of defect replacement | 0.004 | 0.4 |
| Geographic location | 0.046 | 4.6 | Cost escalations | 0.004 | 0.4 |
| Defect replacement | 0.035 | 3.5 | logistics | 0.003 | 0.3 |
| Credit period | 0.035 | 3.5 | Conduct audit | 0.003 | 0.3 |
| Competitive product price | 0.032 | 3.2 | Share info | 0.003 | 0.3 |
| continuous improvement | 0.019 | 1.9 | R & D activities | 0.003 | 0.3 |

The results show values arranged in decreasing order so that prioritization of the sub-criteria can be accomplished (Table 21). It is observed that the quality and the delivery criteria occupy the top 10 in the ranking list of sub-criteria.

**Step 5: Vendor evaluation**
The aim of adopting the AHP technique was to assign weights to the different criteria for evaluating the vendor in XYZ steel pipe manufacturing company. After assigning the weights for the criteria and sub-criteria and validating the model, a rating scale was given to the senior executives of the procurement department. The respondents were asked to rate the raw material vendors on a scale from 0-10; 0 being the worst and ten being the best (Table 22).

Table 22
Ten point Likert scale

| Ratings | 0 | 1 | 2 | 3 | 4 | 5 | 6 | 7 | 8 | 9 | 10 |
|---|---|---|---|---|---|---|---|---|---|---|---|
| Preference | Worst | Very poor | Poor | Significantly below Avg. | Below Avg. | Avg. | Above Avg. | Significantly above Avg. | Good | Very Good | Best |

The ratings for the different vendors for the API and IS vendors of raw materials for XYZ company are given below in Tables 23 and 24 respectively. The sum of global weights of the vendors was computed, and the vendor with the highest total global weights is considered the best vendor based on the criteria and sub-criteria. Vendors A and E are the manufacturers of API specification mild steel coils (raw material for pipe manufacturing) and Vendors B, C and D are the dealers and distributors of both API and IS specification raw materials. Vendors P and Q are the producers of IS specification raw materials. Figure 2 graphically shows the weights of the API vendors multiple criteria that were considered. Similarly, Figure 3 shows the weights of the IS vendors criteria that were considered in this study.

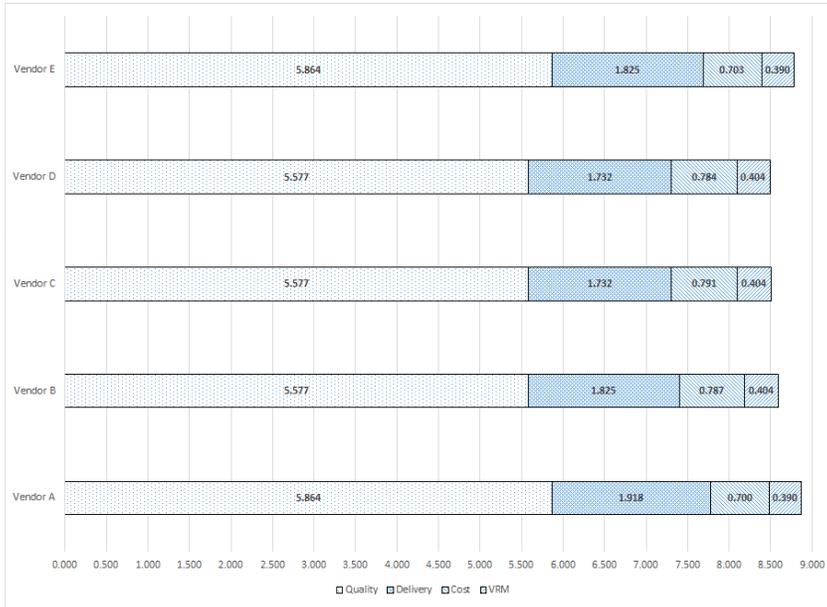

Figure 2. Weights of API Vendors w.r.t criteria

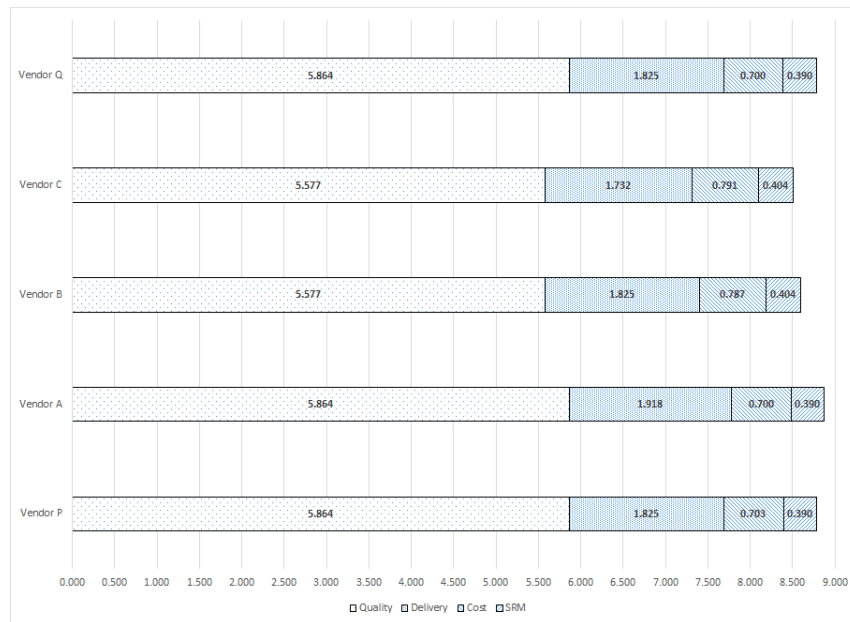

Figure 3. Weights of API Vendors w.r.t criteria

Table 23
API vendor's ratings and weights with respect to different criteria

| Sub criteria | Global weights | Vendor A Ratings | Vendor A Weights | Vendor B Ratings | Vendor B Weights | Vendor C Ratings | Vendor C Weights | Vendor D Ratings | Vendor D Weights | Vendor E Ratings | Vendor E Weights |
|---|---|---|---|---|---|---|---|---|---|---|---|
| S | 0.130 | 9 | 1.166 | 9 | 1.166 | 9 | 1.166 | 9 | 1.166 | 9 | 1.166 |
| C | 0.130 | 9 | 1.166 | 9 | 1.166 | 9 | 1.166 | 9 | 1.166 | 9 | 1.166 |
| TC | 0.130 | 9 | 1.166 | 8 | 1.037 | 8 | 1.037 | 8 | 1.037 | 9 | 1.166 |
| I | 0.124 | 9 | 1.112 | 9 | 1.112 | 9 | 1.112 | 9 | 1.112 | 9 | 1.112 |
| TS | 0.121 | 9 | 1.085 | 8 | 0.964 | 8 | 0.964 | 8 | 0.964 | 9 | 1.085 |
| GC | 0.067 | 9 | 0.599 | 9 | 0.599 | 9 | 0.599 | 9 | 0.599 | 9 | 0.599 |
| F | 0.047 | 9 | 0.421 | 8 | 0.374 | 7 | 0.328 | 7 | 0.328 | 8 | 0.374 |
| GL | 0.046 | 9 | 0.417 | 8 | 0.370 | 7 | 0.324 | 7 | 0.324 | 8 | 0.370 |
| DR | 0.035 | 9 | 0.317 | 9 | 0.317 | 9 | 0.317 | 9 | 0.317 | 9 | 0.317 |
| CP | 0.035 | 8 | 0.276 | 9 | 0.311 | 9 | 0.311 | 9 | 0.311 | 8 | 0.276 |
| CPP | 0.032 | 8 | 0.255 | 9 | 0.286 | 9 | 0.286 | 9 | 0.286 | 8 | 0.255 |
| CI | 0.019 | 9 | 0.168 | 7 | 0.130 | 7 | 0.130 | 7 | 0.130 | 9 | 0.168 |
| LR | 0.018 | 8 | 0.145 | 9 | 0.164 | 9 | 0.164 | 9 | 0.164 | 8 | 0.145 |
| BR | 0.014 | 9 | 0.127 | 9 | 0.127 | 9 | 0.127 | 9 | 0.127 | 9 | 0.127 |
| DM | 0.012 | 9 | 0.109 | 9 | 0.109 | 9 | 0.109 | 9 | 0.109 | 9 | 0.109 |
| BOD | 0.011 | 8 | 0.085 | 9 | 0.096 | 9 | 0.096 | 9 | 0.096 | 8 | 0.085 |
| TI | 0.006 | 9 | 0.056 | 9 | 0.056 | 9 | 0.056 | 9 | 0.056 | 9 | 0.056 |
| SIM | 0.005 | 8 | 0.040 | 9 | 0.045 | 9 | 0.045 | 9 | 0.045 | 8 | 0.040 |
| DRC | 0.004 | 8 | 0.035 | 8 | 0.035 | 8 | 0.035 | 8 | 0.035 | 8 | 0.035 |
| CE | 0.004 | 7 | 0.026 | 9 | 0.033 | 9 | 0.033 | 9 | 0.033 | 7 | 0.026 |
| L | 0.003 | 7 | 0.023 | 8 | 0.027 | 9 | 0.030 | 7 | 0.023 | 8 | 0.027 |
| CA | 0.003 | 9 | 0.030 | 7 | 0.023 | 7 | 0.023 | 7 | 0.023 | 9 | 0.030 |
| SI | 0.003 | 7 | 0.023 | 9 | 0.030 | 9 | 0.030 | 9 | 0.030 | 7 | 0.023 |
| RDA | 0.003 | 8 | 0.025 | 5 | 0.015 | 5 | 0.015 | 5 | 0.015 | 8 | 0.025 |
| Sum | | | 8.872 | | 8.593 | | 8.503 | | 8.496 | | 8.782 |

S-API/IS spec, C-Certified, TC-Test Certificate ,I-Inspection, TS-Technical Specification, GC-Good condition, F-Flexibility, GL-Geographic Location, DR-Defect Replacement, CP-Credit Period, CPP-Competitive Product Price, CI-Continuous Improvement, LR-Long-term Relationship, BR-Background Reputation, DM-Delivery Mode, BOD-Bulk Order Discounts, TI-Tracking Information, SIM-Suggestion Implementation, DRC-Defect Replacement Cost, CE-Cost Escalations, LR-Logistics, CA-Conduct Audit, SI-Share Info, RDA-R & D activities

*IJAHP Article: Kamath, Naik, Prasad/ Vendor's evaluation – using AHP for an Indian steel pipe manufacturing company*Table 24
IS vendor's ratings and weights with respect to different criteria

| Sub criteria | Global weights | Vendor P Ratings | Vendor P Weights | Vendor A Ratings | Vendor A Weights | Vendor B Ratings | Vendor B Weights | Vendor C Ratings | Vendor C Weights | Vendor Q Ratings | Vendor Q Weights |
|---|---|---|---|---|---|---|---|---|---|---|---|
| S | 0.130 | 9 | 1.166 | 9 | 1.166 | 9 | 1.166 | 9 | 1.166 | 9 | 1.166 |
| C | 0.130 | 9 | 1.166 | 9 | 1.166 | 9 | 1.166 | 9 | 1.166 | 9 | 1.166 |
| TC | 0.130 | 9 | 1.166 | 9 | 1.166 | 8 | 1.037 | 8 | 1.037 | 9 | 1.166 |
| I | 0.124 | 9 | 1.112 | 9 | 1.112 | 9 | 1.112 | 9 | 1.112 | 9 | 1.112 |
| TS | 0.121 | 9 | 1.085 | 9 | 1.085 | 8 | 0.964 | 8 | 0.964 | 9 | 1.085 |
| GC | 0.067 | 9 | 0.599 | 9 | 0.599 | 9 | 0.599 | 9 | 0.599 | 9 | 0.599 |
| F | 0.047 | 8 | 0.374 | 9 | 0.421 | 8 | 0.374 | 7 | 0.328 | 8 | 0.374 |
| GL | 0.046 | 8 | 0.370 | 9 | 0.417 | 8 | 0.370 | 7 | 0.324 | 8 | 0.370 |
| DR | 0.035 | 9 | 0.317 | 9 | 0.317 | 9 | 0.317 | 9 | 0.317 | 9 | 0.317 |
| CP | 0.035 | 8 | 0.276 | 8 | 0.276 | 9 | 0.311 | 9 | 0.311 | 8 | 0.276 |
| CPP | 0.032 | 8 | 0.255 | 8 | 0.255 | 9 | 0.286 | 9 | 0.286 | 8 | 0.255 |
| CI | 0.019 | 9 | 0.168 | 9 | 0.168 | 7 | 0.130 | 7 | 0.130 | 9 | 0.168 |
| LR | 0.018 | 8 | 0.145 | 8 | 0.145 | 9 | 0.164 | 9 | 0.164 | 8 | 0.145 |
| BR | 0.014 | 9 | 0.127 | 9 | 0.127 | 9 | 0.127 | 9 | 0.127 | 9 | 0.127 |
| DM | 0.012 | 9 | 0.109 | 9 | 0.109 | 9 | 0.109 | 9 | 0.109 | 9 | 0.109 |
| BOD | 0.011 | 8 | 0.085 | 8 | 0.085 | 9 | 0.096 | 9 | 0.096 | 8 | 0.085 |
| TI | 0.006 | 9 | 0.056 | 9 | 0.056 | 9 | 0.056 | 9 | 0.056 | 9 | 0.056 |
| SIM | 0.005 | 8 | 0.040 | 8 | 0.040 | 9 | 0.045 | 9 | 0.045 | 8 | 0.040 |
| DRC | 0.004 | 8 | 0.035 | 8 | 0.035 | 8 | 0.035 | 8 | 0.035 | 8 | 0.035 |
| C | 0.004 | 7 | 0.026 | 7 | 0.026 | 9 | 0.033 | 9 | 0.033 | 7 | 0.026 |
| L | 0.003 | 8 | 0.027 | 7 | 0.023 | 8 | 0.027 | 9 | 0.030 | 7 | 0.023 |
| CA | 0.003 | 9 | 0.030 | 9 | 0.030 | 7 | 0.023 | 7 | 0.023 | 9 | 0.030 |
| SI | 0.003 | 7 | 0.023 | 7 | 0.023 | 9 | 0.030 | 9 | 0.030 | 7 | 0.023 |
| RDA | 0.003 | 8 | 0.025 | 8 | 0.025 | 5 | 0.015 | 5 | 0.015 | 8 | 0.025 |
| Sum |  |  | 8.782 |  | 8.872 |  | 8.593 |  | 8.503 |  | 8.779 |

*International Journal of the Analytic Hierarchy Process*    457    *Vol. 8 Issue 3 2016*
*ISSN 1936-6744*
*http://dx.doi.org/10.13033/ijahp.v8i3.460*

Table 25
API and IS vendor's ranking

|       | API     |         | IS      |         |
|-------|---------|---------|---------|---------|
| Sl.no | Vendors | Weights | Vendors | Weights |
| 1     | Vendor A | 8.872  | Vendor A | 8.872  |
| 2     | Vendor E | 8.782  | Vendor P | 8.782  |
| 3     | Vendor B | 8.593  | Vendor Q | 8.779  |
| 4     | Vendor C | 8.503  | Vendor B | 8.593  |
| 5     | Vendor D | 8.496  | Vendor C | 8.503  |

Thus, from the above study it can be shown that Vendor A for both API and IS specification have the highest weight.

## 4. Results and discussion

From the above analysis, it can be determined that the quality criterion is the most important factor in vendor evaluation with the preference score of 0.652. It was also found that delivery is the second most important criterion with the score of 0.213 followed by the cost (0.088) and vendor relationship management (0.047). From the scores computed for the various criteria it can be concluded that the quality of the raw material carries much weight when compared to the other criteria with respect to the firm being studied. The quality followed by price, profile of the vendor and service are the most important contributors for selecting a vendor. This proves earlier studies that state that quality is an important criterion for vendor evaluation (Tam & Tummala, 2001). Tahriri, Osman et. al (2008) support the results of this study since they also suggested that quality is an important criterion followed by delivery, cost and responsiveness of the vendors. In order for companies to be sustainable in the competitive market, it is necessary for the firm to provide the best quality product. Therefore, quality is an attribute which is a growing indicator for the success of the firm, and the quality of the raw material is essential for the company to provide the best quality product.

Vendor E should be selected as the best raw material vendor for both API and IS specification (8.872) according to the ranking shown in Table 25. In the case of API specification vendors, it can be noted that the global weights of Vendors A and E quality criterion levels are same (5.864) (Figure 2). Nevertheless, when it came to the delivery criterion Vendor A was ranked highest with the score of 1.918 compared to Vendor E at 1.825. Therefore, Vendor A is given the highest preference. In the case of IS specification vendors, it can be noted that the global weights of Vendors A, P and Q quality criterion levels are same (5.864) (Figure 3). At the same time, when it came to the delivery criterion, Vendor A was ranked highest with the score of 1.918 compared to Vendors P and Q at 1.825. Vendor P was ranked higher than Q when compared to cost criterion. Hence, Vendor A was ranked the highest compared to all the other vendors evaluated based on the overall weights computed. Though the weights of Vendor A were low in VRM when compared to the other vendors, it is still ranked highest as the weights of the quality and delivery criteria are more compared to the other criteria. For XYZ Company, the combined criteria rank Vendor A as the best vendor. The company has to focus more on quality even though its objective is to reduce the cost with the maximization of profit in the supply chain to make sure that their customers are satisfied.

## 5. Conclusion

Every organization should integrate the evaluation and selection process of vendors with the fundamental long term decisions of the firm. The performance of a firm's supply chain activities are directly influenced and the vendor evaluation. Hence, to ensure the maximization of the supply chain activities the process of evaluating, selecting and managing the vendor is very important. The fundamental objective of the process of vendor evaluation is to accomplish world class quality of the product, minimize the cost, speed up delivery times, and attain the best services from the vendor (Tam & Tummala, 2001). With this in mind, the AHP technique was used and applied for the selection of criteria for vendor evaluation for a steel pipe manufacturing company to assess the best API and IS vendors. The mathematical approach given by Saaty (1980) to assign weights for the criteria and sub-criteria was very effective for the evaluation of vendors and selection of the best vendor for the manufacturing company being studied. Areas that need improvement can be determined from the vendor evaluation process. Vendor evaluation helps the organization increase production by ranking vendor qualities. Depending on the various criteria, the alternate vendors can be ranked by the executives of the firm who can make appropriate consideration before ordering the raw materials. Strengths and weaknesses of the various vendors can also be determined by this technique within specific criteria. Firms with a huge expenditure on raw materials need to evaluate vendors and select those who provide excellent value for purchases made.

## 6. Limitations

This section discusses the limitations of the study. First, there is a possibility of response bias because the conclusions of this research were interpreted mainly by Deputy General Managers and managers of the procurement department of XYZ company. Therefore, it is suggested that future research overcome these issues by employing various methodologies such as focus groups, in-depth interviews, and brainstorming sessions with experts etc. Second, the study was conducted specifically for XYZ company, an Indian steel pipe manufacturing industry, and may not be appropriate for other industries or other parts of the world. Third, only a few variables were considered for this study. Other significant factors such as business ethics, government policies, social, political and cultural factors could be included in the future.